\documentstyle[prl,aps,multicol,amsfonts,amstex,epsfig]{revtex}

\renewcommand{\narrowtext}{\begin{multicols}{2} \global\columnwidth20.5pc}
\renewcommand{\widetext}{\end{multicols} \global\columnwidth42.5pc}

\title{The Josephson current in Luttinger liquid-superconductor junctions}

\author {J.-S. Caux$^1$, H. Saleur$^2$ and F. Siano$^2$}

\address{$^1$Theoretical Physics, University of Oxford, 1 Keble Road,
Oxford, OX1 3NP, UK and All Souls College, Oxford OX1 4AL, UK \\
$^2$ Department of Physics and Astronomy, University of Southern
California, Los Angeles, CA 90089} 

\date{$Date:$ \today}

\begin{document}

\bibliographystyle{prsty}

\draft
\maketitle

\begin{abstract}
We study the Josephson current through a Luttinger liquid 
in contact with two superconductors.   We show that
it can be deduced from the Casimir energy in a two-boundary
version of the sine-Gordon model.  We develop a new thermodynamic
Bethe Ansatz, which, combined with a subtle analytic continuation
procedure, allows us to calculate this energy in closed form,
and obtain the complete current-crossover function 
from the case of complete normal to complete Andreev reflection.
\end{abstract}



\narrowtext 
Low-dimensional condensed matter systems exhibit an exciting set of
unusual properties, which have been actively studied theoretically for
some time, and have recently become a topic of explosive experimental
interest.  The physics of these systems is crucially related to 
interactions, which in low dimensions behave nonperturbatively, 
giving rise to such nonintuitive phenomena as spin and charge
separation, or charge fractionalization.  

Dealing with these interactions theoretically is a challenge,
especially as far as transport properties are concerned.  These are of
course the most relevant from an experimental point of view.  New
methods have had to be developed over the years to tackle these
issues, 
like bosonization, conformal field theory and
integrability.  The area where most progress has been made is probably
that of single impurity problems, including the Kondo effect
\cite{TsvelikAP32,AffleckNPB360}, edge
states tunneling \cite{FendleyPRL74}, and quantum dots.

One of the paradigms of low-dimensional electronic systems is the
Luttinger liquid, and it can only be expected that fascinating
properties should appear when combined with another great paradigm of
solid state physics:  superconductivity.  Indeed, it was recently
shown that, as a result of the low-dimensional interactions, junctions
between superconductors and Luttinger liquids behaved very differently
\cite{MaslovPRB53,FazioPRB53,TakaneJPSJ66,AffleckPRB62} from the usual
ones \cite{BlonderPRB25}:  in particular, there is now an RG {\it
flow} as the 
temperature or the length of the junction is changed, leading, for
repulsive (resp. attractive) bulk interactions to perfect normal
(resp. Andreev) reflection at low energies \cite{AffleckPRB62}.  In
the non-interacting 
case there is no flow, and the relative amounts of normal and Andreev
reflections are set by parameters like the superconducting gaps only.

The formalism developed so far allows one to predict global features
of the RG flow, but not to exactly compute physical quantities like
the current.  This is due to the presence of Luttinger interactions,
which require a non perturbative approach.  In this letter, we present
a new formalism to deal with this difficulty, and show how it gives
rise to closed form expressions for the current, all the way from the
UV to the IR fixed points.  

We use here the formulation developed in \cite{AffleckPRB62}.  
Figure 1 illustrates our considerations.  
A quantum wire 
is suspended between two bulk
superconductors $S_{l,r}$.
We describe $S_l$ and $S_r$   
with BCS theory, using gaps $\Delta_l^{(bulk)}$ and $\Delta_r^{(bulk)}
e^{i\chi}$, 
with $\Delta_{l,r}^{(bulk)}$ real numbers.  The phase difference
between the gaps, $\chi$, is crucial in driving the 
Josephson current.
Since the electronic states in 
$S_l$ and $S_r$ are suppressed by the bulk BCS gaps, 
an effective theory at low energies (compared to the gaps) can be
obtained by integrating out 
all states in $S_l$ and $S_r$.  
\begin{figure}
\centerline{\epsfig{figure=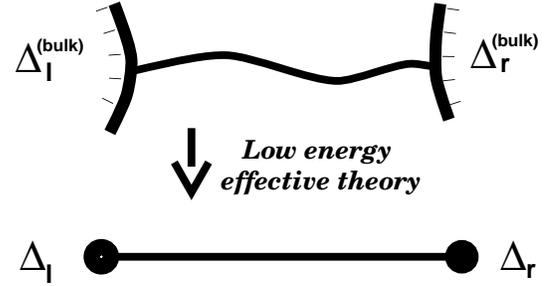,width=7cm}}
\vspace{0.3cm}
\caption{The effective theory for a Josephson junction with a quantum
wire between two
superconductors $S_1$ and $S_2$, with bulk BCS gaps
$\Delta_{1,2}^{(bulk)}$:  a bounded Luttinger liquid, 
with BCS-like couplings $\Delta_{l,r}$ living on the boundaries only.}
\label{Fig.1}
\end{figure}
The only 
contributions left from the original superconductors are two boundary
BCS terms modifying the Luttinger liquid Hamiltonian:  
in terms of fermionic left- and right-movers, we end up with the boundary 
contributions
\begin{eqnarray}
 H_B \propto \Delta_l \Psi^{\dagger}_{R\uparrow}(0)
\Psi^{\dagger}_{L\downarrow}(0) + \Delta_r e^{i \chi} 
\Psi^{\dagger}_{R\uparrow}(R)
\Psi^{\dagger}_{L\downarrow}(R) + ...
\end{eqnarray}
in which $...$ means $h.c. + L \leftrightarrow R$.
The values of the boundary parameters $\Delta_{l,r}$ are
related to the original bulk gaps $\Delta_{l,r}^{(bulk)}$ \cite{AffleckPRB62},
and the physical processes at the boundaries depend on them.
Namely, two things can happen when an electron hits one of the
boundaries:  it can be normal-reflected as an electron, 
or Andreev-reflected as the time-conjugate excitation (a hole),
thereby creating an additional Cooper pair in the condensate.  The
relative amounts of normal versus Andreev reflections at each 
boundary is set by $\Delta$:  for $\Delta =0$, only
normal reflection occurs, while for $\Delta \rightarrow
\infty$, only Andreev reflection occurs.  The Josephson current can
thus be physically understood as excitations being Andreev-reflected
back and forth between the two boundaries, thereby transferring 
charge from one end to the other.

Upon bosonization, the theory becomes $H = H_0 + H_B$ where $H_0$ is
the charge part of the Luttinger liquid Hamiltonian, and the boundary
Hamiltonian 
involves the 
dual charge boson (since we don't consider any nontrivial
spin-scattering processes at the boundaries, the spin and charge
sectors decouple.  We thus consider the charge sector only, 
and drop the spin boson in all further formulas):
\begin{eqnarray}
H_B \propto \Delta_l  \cos{\frac{\beta}{2}\tilde{\phi}_c(0)} + \Delta_r
\cos{(\frac{\beta}{2} \tilde{\phi}_c(R) - \chi)}.
\label{boundaryH}
\end{eqnarray}
In the above notation, the interaction parameter is related to
standard bosonization conventions according to $\beta^2 = (4\pi)^2
R_c^2  =4/g_{\rho} = 2/K_{\rho}$.  The current through the system is
then easy to calculate, being given by the difference between right-
and left-mover densities, which under bosonization becomes
$I(x) = -e \sum_{\sigma} (\Psi^{\dagger}_{R \sigma} \Psi_{R \sigma} -
\Psi^{\dagger}_{L \sigma} \Psi_{L \sigma}) 
\propto -e \partial_x \tilde{\phi}_c$.  Reabsorbing this linear
derivative in the quadratic form of the Hamiltonian then shows that
the zero-temperature current is given by the simple relation
$I(\chi) = 2e \frac{\partial}{\partial \chi} E_0 (\Delta_l, \Delta_r,
\chi)$, where $E_0$ is the ground-state energy of the system.

We show in this letter how the Josephson current can be obtained
exactly using the effective two-boundary model described above.  
The first step, generalizing the proposal of \cite{AffleckPRB62}, 
is to think of the bulk theory as the massless limit of 
a sine-Gordon theory for the dual charge boson $\tilde{\phi}_c$, and
to use the fact that the boundary sine-Gordon terms in
(\ref{boundaryH}) do not destroy the integrability of the sine-Gordon
model \cite{GhoshalIJMPA9}.  Thus, we can derive our results 
from a study of the massless limit of the {\it double-boundary
sine-Gordon model} 
\begin{eqnarray}
S = \int dt \int_0^R dx \frac{1}{2}\left[(\partial_t 
\phi)^2 - (\partial_x \phi)^2 - m \cos \beta \phi \right] - \nonumber \\
- \int dt \left[ \Delta_l \cos \frac{\beta}{2}
\phi(0) + 
\Delta_r \cos \left(\frac{\beta}{2} \phi(R) -  \chi\right) \right]
\label{2bsG}
\end{eqnarray}
As proposed in \cite{FendleyPRL74}, the key role of integrability is
to provide one with a basis of the free boson Hilbert space made of
excitations that scatter in a simple, factorized way, with no particle
production, both in the bulk and at the boundaries.  In contrast with
\cite{FendleyPRL74} and most other work on boundary integrable field
theories, we have however to deal with a problem involving {\it two}
boundaries.  This gives rise to unexpected subtleties.  

In imaginary time, two quantization schemes
are possible \cite{LeClairNPB453}.  In the first one, called 
the $L-$channel, we  
consider an inverse temperature $L \rightarrow \infty$ and write the
partition function as
\begin{eqnarray}
Z = Tr ~e^{-LH_{l,r}}
\end{eqnarray}
where the trace is over eigenstates of the Hamiltonian with
appropriate boundary conditions at the left and right ends.
Alternately, we can permute the definitions of space and imaginary
time and move to the $R-$channel picture, where time flows from one
``boundary'' state to the other.  In this case, the partition function
is given by 
\begin{eqnarray}
Z = \langle B_l | e^{-RH} | B_r \rangle
\label{RchannelZ}
\end{eqnarray}
where the Hamiltonian is quantized along length $L$ with simple
periodic boundary conditions.  Thus, the states $| B_{l,r} \rangle$
become initial and final states, linked through time evolution along
a slice of length $R$.  The advantage of this approach is that
the interpretation of the boundary states  is physically clear:  as no
momentum can flow through the boundaries, only states of zero total
momentum can be emitted or absorbed.  As the boundary sine-Gordon is
an integrable theory \cite{GhoshalIJMPA9}, factorizability of the
scattering imposes very strong constraints on the shape of the
boundary states, constraints which are in fact sufficient to
determine them completely.

We specialize in this paper to the free fermion point
$\beta^2 = 4\pi$, which corresponds to $R_c^2 = 1/4\pi$, in the bulk
attractive regime of the Luttinger liquid.  All difficulties due to
the presence of two boundaries appear in this case, while the bulk
scattering is trivial and does not obscure the issues.  There are
two fundamental excitations in the bulk:  the soliton and
the antisoliton, carrying rapidities $\theta$ (their creation
operators being denoted $A^{\dagger}_{1,2}(\theta)$), with bulk
scattering matrix $S=-1$.  The
boundary scattering however is nontrivial, and involves nondiagonal 
reflection processes.  This affects the boundary states, whose general
form is that of a coherent state of excitations with opposite
rapidities, i.e. \cite{GhoshalIJMPA9}
\begin{eqnarray}
|B_{l,r}> \propto \exp \left\{ \int_{0}^{\infty} d\theta
K^{ab}_{l,r} (\theta) A^{\dagger}_a(-\theta) A^{\dagger}_b(\theta)
\right\} |0> 
\label{boundarystates}
\end{eqnarray}
with amplitudes $K^{ab}_{l,r} (\theta)$ explicitly given in
\cite{AmeduriPLB354}.  

Given the two boundary states,  the partition function can be
calculated in the following way.   For a complete set of states
$\{ \alpha \}$, we have
\begin{eqnarray}
Z = \sum_{\alpha} \frac{\langle B_l | \alpha \rangle \langle \alpha | 
B_r \rangle}{\langle \alpha | \alpha \rangle} e^{-R E_{\alpha}}.
\label{RchannelZagain}
\end{eqnarray}
Now it is clear that the only states having nonzero internal product
with the boundary states are those involving pairs of
excitations:  if rapidity $\theta$ is occupied, so must rapidity
$-\theta$.  As we are going to do thermodynamics using these states
as a basis, it is important to classify them properly.  
Using only positive rapidities for the labeling, we can
construct four different pairs, which we label $11,22,12,21$.  A
pair $12$ has for example the meaning of a soliton at rapidity
$-\theta$ and an antisoliton at rapidity $\theta$.  
A difficulty then occurs when one implements fermionic statistics.
Whereas fermions of the same type cannot occupy the same rapidity
state, we can in fact superimpose some pairs, e.g. pairs $11$
and  $22$ but not $12$ and $22$.  
To do the thermodynamics, it is useful to consider that 
in fact none of these pairs can be superimposed, and then account
for the possible superpositions by introducing one additional species,
a quartet, obtained by superposition of $11$ and $22$, or,
equivalently, $12$ and $21$.  This appears in the boundary 
state with amplitude $K^{11}_{l,r} K^{22}_{l,r} -
K^{12}_{l,r} K^{21}_{l,r} = \hbox{det} (K_{l,r})$.  The trace over
states $\alpha$ now becomes a trace over configurations of these
five excitations on the lattice of allowed rapidities.  The fugacity
of $\alpha$ as determined by the boundary states reads
\begin{eqnarray}
\frac{\langle B_l | \alpha \rangle \langle \alpha | 
B_r \rangle}{\langle \alpha | \alpha \rangle} = \hspace{4cm} \nonumber \\
=\prod_{i=1}^N 
\bar{K}^{b_i a_i}_l (\theta_i) K^{a_i b_i}_r (\theta_i) 
\prod_{j=1}^{N_q} \det [\bar{K}_l (\theta^q_j) K_r(\theta^q_j)]
\end{eqnarray}
where in this example we take $\alpha$ to have $N$ pairs of types
$a_i, b_i$ at 
rapidities $\theta_i, i = 1,...,N$, and $N_q$ quartets at rapidities
$\theta^q_i, i = 1,...,N_q$.  
The TBA can then be done \cite{CauxTBP} according to standard
procedures by looking 
for a saddle point in the sum (\ref{RchannelZagain}), yielding an
expression for $\ln Z$ which we can reinterpret as the
zero-temperature ground-state energy when $L\rightarrow \infty$.  The
end result is
\begin{eqnarray}
E_0 = \lim_{m \rightarrow 0} \frac{-1}{2\pi} \int_0^{\infty} d \theta
\ln \biggl[ 
1 + tr [\bar{K}_l K_r] e^{-2mR \cosh \theta} + \biggr. \nonumber \\
\biggl. + \det [\bar{K}_l K_r]
e^{-4mR \cosh \theta} \biggr] m \cosh \theta.
\end{eqnarray}
Taking the limit, and substituting for the $K$ matrices from
\cite{AmeduriPLB354}, this formula becomes
\begin{eqnarray}
E_0 = \frac{-1}{4\pi} \int_0^{\infty} d \kappa \ln \left[
2\frac{\kappa^2 + 4 \Delta_l^2 
\Delta_r^2 \cos 2\chi}{(\kappa + 2 \Delta_l^2)(\kappa + 2 \Delta_r^2)}
e^{-\kappa R} 
+ \right. \nonumber \\
\left. +
\frac{(\kappa - 2 \Delta_l^2)(\kappa - 2 \Delta_r^2)}{(\kappa + 2
\Delta_l^2)(\kappa 
+ 2 \Delta_r^2)} e^{-2\kappa R} +1 
\right].
\label{groundstate}
\end{eqnarray}
In fact, expression (\ref{groundstate}) hides unexpected subtleties.
To understand the latter, and how to deal with them, it is useful to
discuss for a while a simpler and closely related model, the critical
Ising model with boundary magnetic fields $h_l, h_r$.  As discussed
in \cite{GhoshalIJMPA9}, this theory is integrable, and the same kind
of thermodynamical approach gives rise to the ground-state energy in
a finite size $R$ \cite{CauxTBP}:
\begin{eqnarray}
E_0^{I}(h, h')=-{1\over 4\pi R} \int_0^\infty d\epsilon \times \nonumber \\
\times \ln\left[1+{\epsilon- 8\pi h^2 R\over \epsilon+8\pi h^2 R}
~{\epsilon- 8\pi (h')^2 R\over \epsilon+8\pi (h')^2 R}~
e^{-\epsilon}\right].
\label{TBAi}
\end{eqnarray}
This is in apparent contradiction with physical
expectations.  Indeed, it involves only the {\it squares} of the
magnetic fields, while basic symmetry considerations lead one to
expect an energy that depends on the fields $h_l, h_r$ themselves;
lowest order perturbation theory in fact leads to
\begin{eqnarray}
E_0^{I} (h,h') = \frac{-\pi}{48R} -f(h^2R) - f({h'}^2R) -
2\pi h h' +...
\label{pert}
\end{eqnarray}
with $f(x) = 2 x^2 (cst. + \ln x)/R$.  
The contradiction is partly resolved by realizing that
expression (\ref{TBAi}) expands in fact in the variables
$|h_l|, |h_r|$, and careful manipulations allow one to reproduce
(\ref{pert}) in the case $h_l h_r > 0$.  Expression (\ref{TBAi})
cannot be right however since it does not distinguish (even after the
non-analyticity in $h_{l,r}^2$ has been taken into account) the cases
$h_l h_r > 0$ and $h_l h_r <0$.  These cases should be
different, and we think (\ref{TBAi}) applies to the former only.  A
quick assessment of the situation can be made in the limit of very
large boundary fields, where $h_l h_r >0$ corresponds to $++$ fixed
boundary conditions, where based on conformal field theory we expect
and do indeed get from (\ref{TBAi}) $E_0^I = \frac{-\pi}{48R}$, while
the other case describes $+-$ fixed boundary conditions, for which we
expect $E_0^I = \frac{-\pi}{48R} + \frac{\pi}{2R}$, the difference
being the boundary dimension of the spin operator, $\Delta = 1/2$.

The correct expression for $h_l h_r < 0$ is obtained using the
following physical considerations.  The point is that, for a given
boundary field $h$, the boundary state is fully defined only if one
specifies the boundary conditions asymptotically far away from it.
The results in \cite{GhoshalIJMPA9} assume implicitly that the spin
at infinity is fixed at $+$ while the boundary field is positive.  The
case of a spin fixed $-$ at infinity should be obtained by considering
an 'excited' boundary state.  Following this procedure here leads to
the result that, for $h_l h_r >0$, 
\begin{equation}
E_0^I (h,h')-E_0^I(h,-h')=-{1\over 2} k_+
\label{GSEdiff}
\end{equation}
where $k_+$ is a particular solution of the quantization condition 
$e^{iR k}~
{8\pi h^2+i k\over   8\pi h^2-i k}~ {8\pi
(h')^2+i k\over   8\pi (h')^2-i k}=-1$.  Specifically, all solutions
are analytic functions of $h_l^2$ and $h_r^2$, except
for one pair going as $k_{\pm} \approx \pm 8\pi |h_l h_r|$ for
small fields.  It can be proven that equation (\ref{GSEdiff})
does indeed provide the analytical continuation of (\ref{TBAi}) into the
region $h_l h_r <0$.  For instance, the lowest term in the
perturbative expansion reads now for $h_l h_r <0:  -2\pi |h_l h_r| + 4
\pi |h_l h_r| = 2\pi |h_l h_r| = -2\pi h_l h_r$, as desired.  We also
observe that at large magnetic fields, $k_+ \rightarrow
\frac{\pi}{R}$, producing the expected gap in the conformal limit.
Note that a similar analytical continuation procedure was proposed in
\cite{DoreyNPB525}

\begin{figure}
\centerline{\epsfig{figure=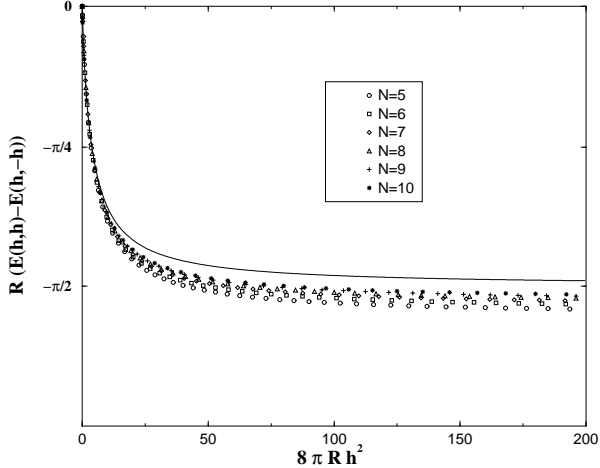,width=6.4cm,angle=270}}
\caption{Ground-state energy difference for the 2-boundary Ising
model with $h_l = \pm h_r$,
for system sizes $N=5$ to $N=10$, and compared to the theoretical
prediction (solid line).}
\label{Fig.2}
\end{figure}

Expression (\ref{GSEdiff}) is extraordinarily simple;  while each of
the terms in the left-hand side requires calculation of a TBA
integral, their difference is simply obtained by the solution of an
elementary quantization condition.  We have checked its validity by
computing numerically the ground state energies of the critical Ising
model with boundary magnetic fields, and studying their scaling
limit.  Results for $h_l = \pm h_r$ will be presented in figure
\ref{Fig.2}, and are in very good
agreement with our formulas.  We note that 
the lattice Ising model with boundary magnetic fields has never been
solved to the best of our knowledge (the continuum version with 
$h_l = h_r$ has been treated in \cite{ChatterjeeMPLA10}), except 
in the case $h_l = 0$ \cite{YangPRB11}, 
where one can check that the results agree with those of our TBA
approach.  

In the sine-Gordon case, we have a similar situation.
Since $E_0$ in (\ref{groundstate}) is an even function
of $\chi$, we can concentrate on the region $\chi \geq 0$.  The
crucial observation is that (\ref{groundstate}) is valid only in the
domain $\chi \in [0, \pi/2[$ since it exhibits a singularity at $\chi
= \pi/2$, where the argument of the log has a zero that hits the real
axis at $\kappa =0$.  Better intuition is gained by studying the
limits $\Delta_{l,r} \rightarrow \infty$, where the integral reduces to
$E_0 = -\frac{1}{4\pi R} \int_0^{\infty} dx \ln \left[ 1 + 2 \cos 2\chi
e^{-x} + e^{-2x} \right]$.
This integral is tabulated, and we get
$E_0 = - \frac{\pi}{24 R} + \frac{\chi^2}{2 \pi R}$ for $\chi
\in [0, \pi/2]$, and
$E_0 = - \frac{\pi}{24 R} + \frac{(\chi - \pi)^2}{2 \pi R}$
for $\chi \in [\pi/2, \pi]$.
The current $I \propto \frac{\partial E}{\partial \chi}$ thus
experiences a discontinuity at $\chi = \pi/2$, which
contradicts perturbation theory:  it should be odd in $\chi$,
and a smooth function over the interval $[0,\pi]$.   As in the Ising
case, our procedure to repair this starts by identifying the
root $k_0$ of the quantization condition
$2\frac{\left[(ik)^2 + g_l^2 g_r^2 \cos 2\chi \right]}{[ik +
g_l^2][ik+g_r^2]} 
e^{-2ikR} + \frac{[ik-g_l^2][ik-g_r^2]}{[ik + g_l^2][ik+g_r^2]}
e^{-4ikR}  = -1$
which requires analytic
continuation beyond $\chi = \pi/2$.  The continuation of the
2-boundary sine-Gordon ground-state energy to all values of $\chi$
then reads
\begin{eqnarray}
E_0^{SG} (\Delta_l,\Delta_r, \chi) &=& E_0, \hspace{2cm} |\chi| <
\pi/2, \nonumber \\ 
E_0^{SG} (\Delta_l,\Delta_r, \chi) &=& E_0 - k_0, \hspace{1cm} \pi/2 <
|\chi| <  \pi. 
\end{eqnarray}
The limiting behaviours of the boundary contributions to the
ground-state energy are $\delta E_0^{SG} \rightarrow
\frac{\chi^2}{2\pi R}$ for
$\Delta_{l,r}$ large, and $\delta E_0^{SG} \propto - \Delta_l \Delta_r
\cos \chi$ for small $\Delta_{l,r}$.

With this formula, we can write down the Josephson current of our
device at the free fermion point $\beta^2 = 4\pi$, in terms of the
dimensionless parameters $\Delta_{L,R}^2 = R \Delta_{l,r}^2$:
\begin{eqnarray}
I (\chi) = \frac{8 e \Delta_L^2 \Delta_R^2}{\pi R} \sin{2\chi}
\int_0^{\infty} d \kappa 
\times \hspace{2cm} \nonumber \\
\times \frac{[\kappa + 2 \Delta_L^2]^{-1} [\kappa + 2\Delta_R^2]^{-1}
e^{-\kappa}} 
{1 + 2 \frac{\kappa^2 + 4 \Delta_L^2 \Delta_R^2 \cos{2\chi}}{(\kappa + 
2\Delta_L^2)(\kappa + 2\Delta_R^2)} e^{-\kappa} + 
\frac{(\kappa - 2\Delta_L^2)(\kappa - 2\Delta_R^2)}{(\kappa + 
2\Delta_L^2)(\kappa + 2\Delta_R^2)} e^{-2\kappa}}, 
\end{eqnarray}
with the understanding that this should be analytically continued
beyond $\chi = \pi/2$ using the above procedure.
This formula for the current provides complete interpolation between
the limits 
of the sawtooth function for perfect Andreev boundary scattering
($\Delta_{l,r} \rightarrow \infty$), and the $\sin{\chi}$ behaviour
for small Andreev scattering.  
These and further results are discussed
in more detail in \cite{CauxTBP}.

J.-S. C. would like to thank J. Cardy, P. Dorey, F. Essler and
A. M. Tsvelik for useful discussions.

\begin{figure}
\centerline{\epsfig{figure=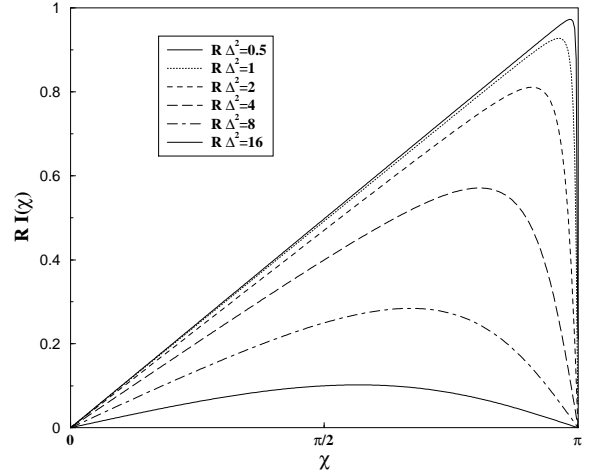,width=6.4cm,angle=270}}
\vspace{0.3cm}
\caption{The Josephson current (in units of $R^{-1}$) as a function of
$\chi$, for $R \Delta^2 = 0.5, 1, 2, 4, 8$ and $16$ (in ascending order).}
\label{Fig.3}
\end{figure}

\widetext

\end{document}